\documentstyle[12 pt]{article}
\def\d{\mbox{\rm d}}
\def\df #1 #2{\displaystyle{\frac{\d #1}{\d #2}}}
\def\dm #1 #2{\displaystyle{\frac{\d^2 #1}{\d #2^2}}}
\def\({\left(}
\def\){\right)}
\def\lib{\left [}
\def\rib{\right]}
\def\re #1{(\ref{#1})}

\def\viz{{\it videlicet}}

\def\dddot#1{\mathinner{\buildrel\vbox{\kern5pt\hbox{...}}\over{#1}}}

\newcommand{\ba}{\begin{array}} 
\newcommand{\ea}{\end{array}} 
\newcommand{\beg}{\begin{eqnarray}} 
\newcommand{\eeq}{\end{eqnarray}} 
\newcommand{\bg}{\begin{eqnarray*}} 
 
\newcommand{\ed}{\end{eqnarray*}} 
\newcommand{\n}{\newline\hfill}

\newcommand{\p}{\partial} 
 
\newcommand{\notlhd}{\lhd\kern-.8em{/}\ } 
\newcommand{\notexist}{\ \exists\kern-.5em{\raise.1em\hbox{/}}\ } 
\newcommand{\lam}{\Lambda}

\newcommand{\pde}[2]{\frac{\p #1}{\p #2}}

\newcommand{\inp}{{\mbox{\vbox{\hrule width0ex\hbox{\vrule 
 height0ex\kern3.8pt 
\vbox{\kern2.5pt}\kern3.8pt \vrule height1.6ex} 
\hrule width1.6ex}}}}

\begin{document}

\begin{center}
{\large {\bf Linearisable Third Order Ordinary }}
\end{center}
\begin{center}
{\large {\bf Differential Equations}} 
\end{center}
\begin{center}
{\large {\bf and Generalised Sundman Transformations}}
\\[19 mm]
{N Euler, T Wolf\footnote{permanent address: Department of Mathematics,
Brock University, 500 Glenridge Avenue, St. Catharines, Ontario,
Canada L2S 3A1}, PGL Leach\footnote{permanent address: School of
Mathematical and Statistical Sciences, University of Natal, Durban
4041, South Africa} and M Euler}
\\[3 mm]
Department of Mathematics, Lulea University of Technology\\
SE-971 87 Lulea, Sweden
\vspace{7mm}
\end{center}
\strut\vfill

\nonumber
{\bf Copyright \copyright\ 2002 by N Euler, T Wolf, PGL Leach and M Euler}

\pagebreak

\noindent
{\bf N. Euler, T Wolf, PGL Leach and M. Euler}, {\it
 Linearisable Third Order Ordinary Differential Equations
and Generalised Sundman Transformations},
Lulea University of Technology,
Department of Mathematics\newline
Research Report 2 (2002).

\vspace{1cm}

\noindent
\begin{minipage}{350pt}{\bf Abstract}:

\noindent
We calculate in detail the conditions which allow the most
general third order 
ordinary differential
equation to be linearised in $X'''(T)=0$ under the
transformation $X(T)=F(x,t),\ dT=G(x,t)dt$. Further generalisations
are considered.

\vspace{1cm}

\noindent
{\bf Subject Classification (AMS 2000):} 34A05, 34A25, 34A34.

\vspace{1cm}


\vspace{1cm}

\noindent
{\bf Note:} This report has been submitted for publication elsewhere.

\vspace{4cm}

\noindent
{\bf ISSN:} 1400--4003

\vspace{1cm}

\noindent
Lulea University of Technology\n
Department of Mathematics\n
S-97187 Lulea, SWEDEN

\pagebreak

\end{minipage}
\vspace{7mm}

\strut\vfill

\pagebreak

\section{Introduction}

In the modelling of physical and other phenomena differential
equations, be they ordinary or partial, scalar or a system, are a common
outcome of the modelling process.  The basic problem becomes the
solution of these differential equations.  One of the fundamental
methods of solution relies upon the transformation of a given equation
(or equations; hereinafter the singular will be taken to include the
plural where appropriate) to another equation of standard form.  The
transformation may be to an equation of like order or of greater or
lesser order.  In the early days of the solution of differential
equations at the beginning of the eighteenth century the methods for
determining suitable transformations were developed very much on an
{\it ad hoc} basis.  With the development of symmetry methods,
initiated by Lie towards the end of the nineteenth century and
revived as well as developed ever since, the {\it ad hoc} methods
were replaced by systematic approaches.  About the same time classes
of equations were established as equivalent to certain
standard equations.  In particular the possibility that a
given equation could be linearised, {\it i.e.}\ transformed to a linear
equation, was a most attractive proposition due to the special
properties of linear differential equations.  Already in his thesis of 1896
Tresse \cite{Tresse 96} showed that the most general second order ordinary differential equation which could be transformed to the simplest second order equation, \viz
\begin{equation}
X''(T) = 0,\label{1.1}
\end{equation}
by means of the point transformation
\begin{equation}
X = F (x,t)\quad T = G (x,t)\label{1.2}
\end{equation}
has the form
\begin{equation}
\ddot{x} +\Lambda_3(x,t)\dot{x}^3+\Lambda_2(x,t)\dot{x}^2
+\Lambda_1(x,t)\dot{x} +
\Lambda_0(x,t) = 0,\label{1.3}
\end{equation}
where the overdot denotes differentiation with respect to
the independent variable $t$ whereas the primes refer to $T$
derivatives. This notation is used throughout the paper.
In terms of the transformation
functions $F$ and $G$ the functions $\Lambda_i$
are given by
\begin{eqnarray}
\Lambda_3(x,t) & = &\lib F_{xx}G_x-F_xG_{xx}\rib/J\nonumber\\
\Lambda_2(x,t) & = &\lib F_{xx}G_t+ 2F_{xt}G_x- 2F_xG_{xt} -F_tG_{xx}\rib/J\nonumber\\
\Lambda_1(x,t) & = &\lib F_{tt}G_x+ 2F_{xt}G_t- 2F_tG_{xt} -F_xG_{tt}\rib/J\nonumber\\
\Lambda_0(x,t) & = &\lib F_{tt}G_t-F_tG_{tt}\rib/J,\label{1.6}
\end{eqnarray}
where $J(x,t) =F_xG_t-F_tG_x\neq 0 $ is the Jacobian of the point
transformation \re{1.2}.  As usual subscripts denote partial derivatives.
Through the elimination of the transformation
functions
$F $ and $G $ it is found that the coefficient functions,
$\Lambda_i,\ i = 0,\ldots, 3, $
must satisfy the conditions
\begin{eqnarray}
&&\Lambda_{1xx}-2\Lambda_{2xt} + 3\Lambda_{3tt} + 6\Lambda_3\Lambda_{0x} +
3\Lambda_0\Lambda_{3x} - 3\Lambda_3\Lambda_{1t} - 3\Lambda_1\Lambda_{3t}
\nonumber\\ & & \quad-\Lambda_2\Lambda_{1x} + 2\Lambda_2\Lambda_{2t}  =  0\label{1.4}\\
&&\Lambda_{2tt} - 2\Lambda_{1xt} + 3\Lambda_{0xx} - 6\Lambda_0\Lambda_{3t} -
3\Lambda_3\Lambda_{0t} + 3\Lambda_0\Lambda_{2x} + 3\Lambda_2\Lambda_{0x}
\nonumber\\ & & \quad+\Lambda_1\Lambda_{2t} - 2\Lambda_1\Lambda_{1x}  =  0.\label{1.5}
\end{eqnarray}
The problem has attracted some interest since the work of Tresse.
See \cite{Duarte 87, Sarlet 87, Duarte 94, Euler 97}.
Note that the condition
corresponding to \re{1.4} given in \cite{Duarte 94} omits the
coefficient 2 in the final term.

In a practical context one identifies the coefficient functions from
the given nonlinear equation and substitutes them in the compatibility
conditions \re{1.4} and \re{1.5} to determine whether the equation is
of the correct form.  If this be the case, the transformation is
determined from the solution of the system \re{1.6} and the solution
to the nonlinear equation follows immediately.  Since all second order
linear ordinary differential equations are equivalent to \re{1.1}
under a point transformation ({\it i.e.}\ of the type \re{1.2}), these
formul\ae\ answer the question of the class of second order equations linearisable under a point transformation.

The attraction of point transformations is that they preserve Lie point
symmetries.  Since a second order linear equation possesses eight Lie
point symmetries, which is far in excess of the two required to reduce
the equation to quadratures, there is no necessity to confine one's
interest to point transformations if the matter of interest is the
solution of the equation and not its point symmetries.  One can then look
towards some type of transformation which is more general than a point
transformation.  The advantage of point transformations is that they
are fairly easy to work with.  In looking towards a
generalisation one wants to keep this property, if possible.

A convenient form of generalisation is the nonpoint
transformation introduced by Duarte {\it et al}\ \cite{Duarte 94}
which has the form
\begin{equation}
X(T) =F (t,x),\qquad d T = G (t,x)d t.\label{1.7}
\end{equation}
Without a knowledge of the functional form of $x (t) $ the
transformation \re{1.7} is a particular type of nonlocal
transformation.
This transformation is a generalisation of a transformation proposed
by Sundman \cite{Sundman 12}.  Since the expression 'non-point
transformation' has a meaning more general than that of \re{1.7},
it may be better to refer to this type of transformation as a
{\bf generalised Sundman transformation}.

In their paper \cite{Duarte 94} Duarte, Moreira and Santos
derived the most general conditions for
which a second order ordinary differential equation may be
transformed to the free particle equation $X''=0$ under the
generalised Sundman transformation (\ref{1.7}).
Euler \cite{Euler 97} studied the general anharmonic oscillator
\beg
\label{anharm}
\ddot x+f_1(t)\dot x+f_2(t)x+f_3(t)x^n=0
\eeq
and derived conditions
on the coefficient functions $f_j$ for which (\ref{anharm})
may be linearised under the generalised Sundman transformation
(\ref{1.7}). It follows \cite{Euler 97} that (\ref{anharm})
may be reduced to the linear equation
\beg
\label{lin_2nd}
X''+k=0,\qquad k\in \Re\backslash\{0\}
\eeq
by the transformation
\beg
\label{trans_2nd}
X(T)=h(t)x^{n+1},\qquad dT=\left(\frac{n+1}{k}f_3(t)h(t)\right)\ x^n\,dt,
\eeq
where $n\in {\cal Q}\backslash\{-3,-2,-1,0,1\}$ and
\beg
\label{h_lin}
h(t)=f_3^{(n+1)/(n+3)}\exp\left\{2\left(\frac{n+1}{n+3}\right)\int^tf_1(\rho)
d\rho\right\},
\eeq
if and only if $f_1,\ f_2$ and $f_3$ satisfy the following condition:
\beg
\label{fs_con}
& &f_2=\frac{1}{n+3}\frac{\ddot f_3}{f_3}
-\frac{n+4}{(n+3)^2}\left(\frac{\dot f_3}{f_3}\right)^2
+\frac{n-1}{(n+3)^2}\left(\frac{\dot f_3}{f_3}\right)f_1\nonumber\\
& & \nonumber\\
& & \qquad +2\frac{1}{n+3}\dot f_1 +2\frac{n+1}{(n+3)^2}f_1^2.
\eeq
This leads to the invariant (time-dependent first integral) of
(\ref{anharm}) through the first integral of (\ref{lin_2nd}), which is
\bg
I(X,X')=X+\frac{1}{2k}\left(X'\right)^2.
\ed

Within the two classes of transformation given by \re{1.2} and
\re{1.7} there are subclasses which have some particular interest if
one is concerned about the type of transformation and preservation of
types of symmetry.  For example in the case that $G_x = 0 $ both
classes of transformations reduce to a transformation of
Kummer-Liouville type \cite{Kummer 34, Liouville 37} which has the
property of preserving symmetries of Cartan form, {\it i.e.}\ fibre-preserving
transformations \cite{Kamran 86}. These transformations are of
importance for Hamiltonian Mechanics and Quantum Mechanics.

In the studies of second order equations use was made of the fact
that every linear second order equation is equivalent under a point
transformation to \re{1.2}.  We recall \cite{Lie 70} [p 405] that the
number of Lie point symmetries of a second order equation can be
0, 1, 2, 3 or 8 and that all linear second order equations have
eight Lie point symmetries with the Lie algebra $sl (2,R) $.
Naturally any second order equation linearisable under a point
transformation also has eight Lie point symmetries.  In the case
of third and higher order equations such economy of property does
not persist.  An $n $-th order linear differential equation can
have $n+ 1 $, $n+ 2 $ or $n+ 4 $ Lie point symmetries \cite{Mahomed
90}.
Consequently, even under point transformations, there are three
equivalence classes of linearisable $n $-th order equations.
In addition there are the classes of equations corresponding to
other numbers of Lie point symmetries or different algebras.


The manipulations required for the calculations of the classes of
equations which are equivalent under either of these classes of
transformation are nontrivial.  Although, in principle, the same
ideas -- indeed more complicated types of transformation -- can
be applied to differential equations of all orders, the burden
of calculation has in the past made the widespread use of these
methods impracticable.  These days with the ready availability
of symbolic manipulation codes on personal computers of reasonable
computational power the drudgery of these calculations has been
removed or, at least, transferred to a third nonvocal party.
The time has come to consider the possibilities of 
more complicated transformations of more complicated equations.
In this paper we report the
results for generalised Sundman
transformations
for third order ordinary differential equations.
In our calculations we make use of the packages {\sc Crack}
written in the computer algebra system {\sc Reduce} and
{\sc Rif} written in {\sc Maple}. They are described
in \cite{Wolf 96} and \cite{Reid 96}. 

We concentrate on the basic third order equation
\begin{equation}
X'''(T) = 0,\label{1.12}
\end{equation}
but there is no reason why the ideas presented here cannot be extended
to the other types of third order equation which constitute the set of
equivalence classes of third order equations \cite{Mahomed 88}. We
give only some
examples of more general linear third order equations (see Section 3).
Note
that (\ref{1.12}) admits the two first integrals
\beg
\label{2_first_int}
I_1=X'',\qquad I_2=X'' X-\frac{1}{2}\left(X'\right)^2.
\eeq
The generalised Sundman transformation (\ref{1.7}) can then be used
to express these first integrals to obtain the invariants of the
nonlinear equation derived by the transformation. 

In the case of scalar third order equations there can be 0, 1, 2, 3,
4, 5, 6 and 7 Lie point symmetries.  When the maximum number of Lie
point symmetries is found in an equation, this equation has in
addition three irreducible contact symmetries.
The ten symmetries possess the Lie
algebra $sp (5) $ \cite{Abraham 95}.  The equation \re{1.12} is a
representative of this last class.

We recall that the most general linear differential equation of
third order is
\begin{equation}
\dddot{x} +f_4 (t)\ddot{x} +f_3 (t)\dot{x} +f_2 (t)x+ f_1 (t) = 0.\label{2.1}
\end{equation}
The condition for \re{2.1} to be reduced to
\begin{equation}
X'''(T) = 0\label{2.2}
\end{equation}
by an invertible point transformation was found by Laguerre in 1898
\cite{Laguerre 79} to be
\begin{equation}
\frac{1}{6}\frac{d^2f_4}{dt^2} +\frac{1}{3}f_4\frac{df_4}{dt}
-\frac{1}{2}\frac{df_3}{dt} +\frac{2}{27} f_4^3-\frac{1}{3}f_4f_3+f_2 = 0.\label{2.3}
\end{equation}
We note in passing that it is possible to transform \re{2.1} to
\re{2.2} without any conditions on the coefficient functions if
contact transformations are used either directly 
or as an equivalent point transformation of the corresponding linear
first-order system \cite{Nucci 01a}.


The article is organised as follows: In Section 2 we
present the classes of equation equivalent to \re{1.12} under
the generalised Sundman transformation (\ref{1.7}).
In Section 3 we consider a special Sundman transformation and
show how linearisable ordinary differential equations of second,
third and fourth order and their
invariants can be constructed.
Section 4 is devoted to an extension of the
generalised Sundman transformation, which is related to the so-called
generalised hodograph transformation introduced recently for
evolution equations \cite{Euler NM 01}.
We note that there has been
work done recently on the problem of
transitive fibre-preserving point symmetries of third order
ordinary differential equations \cite{Grebot 97},
as well as contact transformations and local
reducibility of an ordinary differential equation to the form
\re{1.12} \cite{Gusyatnikova 99}.  The results presented here are
complementary to the results presented in those two papers.

\section{Generalised Sundman transformations for $X'''=0$ }
\setcounter{equation}{0}

We turn our attention to the equivalence class of third order
nonlinear ordinary differential equations obtainable from \re{2.2},
\viz
\bg
X'''(T)=0
\ed
by
means of the
generalised Sundman transformation \re{1.7}, \viz
\bg
X(T)=F(t,x),\qquad dT=G(t,x)dt,
\ed
where $F,\ G\in {\cal C}^{3}$ and are to be determined for the
transformation
of (\ref{2.2}).
Provided $F_xG^2\neq 0 $, the form of the representative equation of the equivalence class is
\begin{equation}
\dddot{x} +\Lambda_5(x,t)\ddot{x} +\Lambda_4(x,t)\dot{ x}\ddot{x}
+\Lambda_3(x,t)\dot{ x}^3+\Lambda_2(x,t)\dot{x}^2
+\Lambda_1(x,t)\dot{x}
+\Lambda_0(x,t) = 0,\label{3.1}
\end{equation}
where the functions $\Lambda_i$ are related to the transformation
functions $F$ and $G$ by means of
\begin{eqnarray}
\Lambda_5(x,t) & = &3\frac{F_{xt}}{F_x} - 3\frac{G_t}{G} -\frac{F_t}{F_x}\frac{G_x}{G}\nonumber\\
\Lambda_4(x,t) & = & - 4\frac{G_x}{G} + 3\frac{F_{xx}}{F_x}\nonumber\\
\Lambda_3(x,t) & = &\frac{F_{xxx}}{F_{x}} + 3\(\frac{G_x}{G}\)^2-3\frac{F_{xx}}{F_x}\frac{G_x}{G}\frac{G_{xx}}{G}\nonumber\\
\Lambda_2(x,t) & = & 3\frac{F_{xxt}}{F_x} +
3\frac{F_t}{F_x}\(\frac{G_x}{G}\)^2-6\frac{F_{xt}}{F_x}\frac{G_x}{G} - 3\frac{F_{xx}}{F_x}\frac{G_t}{G}\nonumber\\
& &+ 6\frac{G_x}{G}\frac{G_t}{G} -\frac{F_t}{F_x}\frac{G_{xx}}{G} - 2\frac{G_{xt}}{G}\label{3.2}\\
\Lambda_1(x,t) & = & 3\(\frac{G_t}{G}\)^2+ 3\frac{F_{xtt}}{F_x} -
3\frac{F_{tt}}{F_x}\frac{G_x}{G} - 6\frac{F_{xt}}{F_x}\frac{G_t}{G}
+ 6\frac{F_t}{F_x}\frac{G_x}{G}\frac{G_t}{G}\nonumber \\
& & - 2\frac{F_t}{F_x}
\frac{G_{xt}}{G} -\frac{G_{tt}}{G}\nonumber\\
\Lambda_0(x,t) & = & -\frac{F_t}{F_x}\frac{G_{tt}}{G} - 3\frac{F_{tt}}{F_x}\frac{G_t}{G} +\frac{F_{ttt}}{F_x} + 3\frac{F_t}{F_x}\(\frac{G_t}{G}\)^2.\nonumber
\end{eqnarray}
In order to have the inverse transformation we must establish the compatibility conditions for \re{3.2}.  This is the major thrust of our code.

The above set of equations has the form
\begin{eqnarray}
0 & = E_1 := & 3F_{tx}G-3F_xG_t-F_tG_x-F_xG\Lambda_5              \label{n1} \\
0 & = E_2 := & 3F_{xx}G - 4F_xG_x - F_x\Lambda_4                  \label{n2} \\
0 & = E_3 := & F_{xxx}G^2 - 3F_{xx}G_xG - F_xG_{xx}G
+ 3F_xG_x^{2}- F_xG^2\Lambda_3                                 \label{n3} \\
0 & = E_4 := & 3F_{txx}G^2 - 6F_{tx}G_xG - F_tG_{xx}G +
               3F_tG_x^{2} - 3F_{xx}G_tG \nonumber \\ 
  &          & - 2F_xG_{tx}G + 6F_xG_tG_x - F_xG^2\Lambda_2       \label{n4} \\
0 & = E_5 := & -3F_{ttx}G^2 + 6F_{tx}G_tG + 3F_{tt}G_xG + 2F_tG_{tx}G
               - 6F_tG_tG_x \nonumber \\
  &          & + F_xG_{tt}G - 3F_xG_t^2 + F_xG^2\Lambda_1 \label{n5} \\
0 & = E_6 := & F_{ttt}G^2 - 3F_{tt}G_tG - F_tG_{tt}G 
               + 3F_tG_t^2 - F_xG^2\Lambda_0             \label{n6} 
\end{eqnarray}
In the following simplifications of this  system the
equation to be replaced in each step is multiplied with a
nonvanishing factor so that the new system after each replacement is
still necessary and sufficient. 
\begin{eqnarray*}
E_3 & \rightarrow & E_7 := E_{2x}G - G_x E_2 - 3 E_3 \\
E_7 & \rightarrow & E_8 := \left((5G_x-G \Lambda_4)E_2 - 3E_7\right)/F_x \\
E_4 & \rightarrow & E_9 := E_{2t}G - G_tE_2 - E_4 \\
E_9 & \rightarrow & E_{10} := - G_tE_2 + E_9 \\
E_{10} & \rightarrow & E_{11} := (G\Lambda_4-2G_x)E_1 + 3E_{10} \\
E_5 & \rightarrow & E_{12} := GE_{1t} - G_t E_1 + E_5 \\
E_{12} & \rightarrow & E_{13} := E_1(G\Lambda_5-3G_t) + 3E_{12} \\
E_{11} & \rightarrow & E_{14}:= F_tE_8 - E_{11}
\end{eqnarray*}
The new system consists of equations $E_1,E_2,E_6,E_8,E_{13},E_{14}$ 
and is transformed to two new functions $h$ and $p$ which are 
related to $F$ and $G$ through the relations
\begin{eqnarray}
F(x,t) & = & p(x,t)h^{-1}(x,t)     \label{fsub}\\
G(x,t) & = & h^{-3/2}(x,t).     \label{gsub}
\end{eqnarray}
After making all equations free of a denominator and performing three
simplifications
\begin{eqnarray*}
E_{13} & \rightarrow & E_{15} := (E_{13} - 3h_tE_1)/h \\   
E_{14} & \rightarrow & E_{16} := (E_{14} - 3h_xE_1)/h \\   
E_2    & \rightarrow & E_{17} := (pE_8 - 3E_2)/h
\end{eqnarray*}
we introduce new functions $\Lambda_6, \Lambda_7, \Lambda_8$ through
\begin{eqnarray}
\Lambda_6(x,t) & = & -6\Lambda_{5t}+6\Lambda_1-2\Lambda_5^{2}  \label{l1sub} \\
\Lambda_7(x,t) & = & \;\;\;6\Lambda_{4t} - 6\Lambda_2 + 2\Lambda_4\Lambda_5
\label{l2sub} \\
\Lambda_8(x,t) & = & -6\Lambda_{4x}+18\Lambda_3-2\Lambda_4^{2}. \label{l3sub}
\end{eqnarray}
To compactify the display of the resulting system we use the notation 
$$[A]_{p\leftrightarrow h} := A - A|_{p\leftrightarrow h}$$
where $A$ is a differential expression in the functions $p$ and
$h$ and $A|_{p\leftrightarrow h}$ is the expression after
swapping $p$ and $h$:
\begin{eqnarray}
0 & = & E_8 = 9h_{xx}-3h_x\Lambda_4+h\Lambda_8      \label{s1} \\
0 & = & E_{17} = 9p_{xx}-3p_x\Lambda_4+p\Lambda_8   \label{s2} \\
0 & = & E_6 = [ 2h_{ttt}p + 3h_{tt}p_t - 2h_xp\Lambda_0 ]_{p\leftrightarrow h}  \label{s3} \\
0 & = & E_1 = [ 6h_{tx}p - 3h_tp_x
- 2h_xp\Lambda_5 ]_{p\leftrightarrow h}  \label{s4} \\
0 & = & E_{16} = [ 18h_{tx}p_x - h_tp\Lambda_8
                   + h_xp\Lambda_7 ]_{p\leftrightarrow h} \label{s5} \\
0 & = & E_{15} = [ 9h_{tx}p_t - 18h_{tt}p_x + 3h_tp_x\Lambda_5
                   + h_xp\Lambda_6 ]_{p\leftrightarrow h}. \label{s6}
\end{eqnarray}
Remarkably the system is symmetric under the exchange $p
\leftrightarrow h$.

\strut\hfill

\noindent
{\it Remark:} {\it
In view of the symmetry $p \leftrightarrow h$ and the relations
(\ref{fsub}) and (\ref{gsub}) we obtain the transformation coefficients 
\bg
& & \bar F(x,t)=F^{-1}(x,t)\\
& & \bar G(x,t)=F^{-3/2}(x,t)G(x,t)
\ed
for the generalised Sundman transformation (\ref{1.7}). This does not
lead to new linearisable third order ordinary differential equations,
so we do not list here any conditions for this transformation.}

\strut\hfill

We obtain a first integrability condition by differentiating
equation (\ref{s4})
with respect to $x$ and substituting $h_{txx},h_{xx}$ using 
equation (\ref{s1}),
substituting $p_{txx},p_{xx}$ using equation (\ref{s2}),
substituting $h_{tx}p_x$ using equation (\ref{s5}) and 
substituting $h_{tx}p$ using equations (\ref{s4}). The result is
\begin{equation}
0 = E_{18} := [ (h_x(12\Lambda_{4t}-12\Lambda_{5x}-\Lambda_7) 
                - h_t\Lambda_8)p ]_{p\leftrightarrow h} .       \label{i0}
\end{equation}
Before we give the most general solution we consider two special cases
in the next two subsections.

\subsection{The case $G_x=0$,} {\it i.e.} the transformation
\bg
X(T)=F(x,t),\qquad
dT=G(t)dt,
\ed
where $F(x,t)=p(x,t)h^{-1}(t)$ and $G(t)=h^{-3/2}(t)$.
By investigating the case $G_x=0$ (which is equivalent to $h_x=0$)
we cover the case $p_x=0$
as well because of the $p \leftrightarrow h$ symmetry.

For $h_x=0$ we have $h_t \neq 0$ and get $E_8=0=\Lambda_8$ and further
$E_{16}=0=\Lambda_7$. 
After substitution of $p_{tx}$ from equation (\ref{s4}) into equation (\ref{s6})
the resulting system is
\begin{eqnarray}
0 & = & -E_{18}/(12p_x) =\Lambda_{4t} - \Lambda_{5x}  \label{m1} \\
0 & = & E_{19} := E_{17}/3 = 3p_{xx}-p_x\Lambda_4 \label{m2} \\
0 & = & E_1 = -6hp_{tx}-3h_tp_x+2hp_x\Lambda_5  \label{m3} \\
0 & = & E_{20} := (3h_tE_1-2hE_{15})/p_x = 
          36h_{tt}h - 9h_t^{2} + 2h^2\Lambda_6  \label{m4}  \\
0 & = & E_6 = 2h_{ttt}p +3h_{tt}p_t-3h_tp_{tt}-2hp_{ttt}
+2hp_x\Lambda_0. \label{m5} 
\end{eqnarray}
For equation (\ref{m4}) to have a solution for $h(t)$ the condition on
$\Lambda_6$ is
\begin{equation}
\Lambda_{6x}=0 .       \label{i1}
\end{equation}
We need to derive one more integrability condition before being
able to formulate a procedure to solve the above system. We reduce the condition
\begin{equation}
E_{6x}=0        \label{i2}
\end{equation}
by
substituting $p_{tttx},p_{ttx},p_{tx}$ computed from equation (\ref{m3}),
substituting $p_{xx}$ from equation (\ref{m2}) and 
substituting $h_{tt}$ from equation (\ref{m4}) to get
\begin{eqnarray}
& & 0  =  (108h^2E_{6x}
        - 126hh_{tt}E_1 + 54h_t^{2}E_1 - 36hh_tE_{1t} 
        + 18h_tp_xE_{20} \nonumber \\
&&\qquad  - 6hh_tE_1\Lambda_5 + 36h^2E_{1tt} + 12h^2E_{1t}\Lambda_5 
        - 9hp_xE_{20t}+ 24h^2E_1\Lambda_{5t}  \nonumber \\
&&\qquad  - 6hp_xE_{20}\Lambda_5 
        - 72h^3E_{19}\Lambda_0 + 4h^2E_1\Lambda_5^2)/(2h^3p_x) \label{reduc} \\
&&\quad  =108\Lambda_{0x} - 36\Lambda_{5tt} - 36\Lambda_{5t}\Lambda_5
    - 9\Lambda_{6t} + 36\Lambda_0\Lambda_4 \nonumber\\
& &\qquad - 4\Lambda_5^{3} - 6\Lambda_5\Lambda_6. \label{i3}
\end{eqnarray}

\strut\hfill

\noindent
{\bf We summarize:}
The procedure for a given set of expressions $\Lambda_0, \Lambda_1,
\ldots, \Lambda_5$ is as follows.
\begin{enumerate}
\item Compute $\Lambda_6, \Lambda_7, \Lambda_8$ from equations
      (\ref{l1sub}), (\ref{l2sub}), (\ref{l3sub}).
\item The following set of conditions for $\Lambda_i$ is necessary
      and, as becomes clear below, also
      sufficient for a solution with $h_x=0=G_x$ to exist:
     \[\Lambda_7=0, \;\;\; \Lambda_8=0, \;\;\; \Lambda_{4t} - \Lambda_{5x}=0,
     \;\;\; \Lambda_{6x}=0,\]
     \[108\Lambda_{0x} - 36\Lambda_{5tt} - 36\Lambda_{5t}\Lambda_5
         - 9\Lambda_{6t} + 36\Lambda_0\Lambda_4 - 4\Lambda_5^{3} 
         - 6\Lambda_5\Lambda_6=0. \]
\item The function $h(t)$ is to be computed from the condition
      (\ref{m4}) which is an ordinary differential equation
due to $\Lambda_{6x}=0$ and
      which can be written as a linear equation for $h^{3/4}$:
      \[24(h^{3/4})_{tt} + h^{3/4}\Lambda_6=0.  \] 
\item Compute a function $q(x,t) (=\log(p_x))$ from the two equations
      (\ref{m2}) and (\ref{m3}) as a line integral:
      \[q_x = \frac{1}{3} \Lambda_4, \;\;\; q_t = \frac{1}{3}\Lambda_5
-\frac{1}{2}\frac{h_t}{h}.\]
      The existence of $q$ is guaranteed through condition (\ref{m1}).
\item Compute a function $r(x,t)$ from
      \[r(x,t) = \int\,\exp\left[q(x,t)\right]\,dx.\]
\item The function $p(x,t)$ is computed from
      \[p(x,t) = r(x,t) + s(t),\]
      where the function $s(t)$ is computed from equation (\ref{m5})
      which after the substitution $p=r+s$ takes the form
      \[2h_{ttt}(r+s)+3h_{tt}(r+s)_t-3h_t(r+s)_{tt}-2h(r+s)_{ttt}
+2hr_x\Lambda_0=0.\]
      This is a linear third order equation for $s(t)$.
      It is an ordinary differential equation with a solution for $s(t)$ because the
      derivative of the left hand side of this equation
with respect to $x$ vanishes 
      due to equations (\ref{reduc}), (\ref{m2}), (\ref{m3}), (\ref{m4}) 
      and (\ref{i3}).
\item With the last step we satisfied all conditions (\ref{m1}) -
      (\ref{m5}) and computed $p$ and $h$ which gives $G$ and $F$ through equations
      (\ref{gsub}) and (\ref{fsub}):
      \[G(t) = h^{-3/2}, \;\;\; F(x,t) = ph^{-1}.\]
\end{enumerate}

\subsection{The case $F_t=0$,} {\it i.e.} the transformation
\bg
X(T)=F(x),\qquad dT=G(x,t)dt.
\ed
For all investigations below we can assume 
$h_x \neq 0, p_x \neq 0$. We consider 
$f_t = (p/h)_t = 0$ so that
\begin{equation}
h_t = h\frac{p_t}{p}.   \label{ht}
\end{equation}
A straightforward substitution of $h_t$ from equation (\ref{ht})
into equation (\ref{s4}) gives
\begin{equation}
9p_t - 2p\Lambda_5=0.                                \label{f1} \\  
\end{equation}
Application of this condition on equation (\ref{ht}) yields
\begin{equation}
9h_t - 2h\Lambda_5=0                                \label{f2} \\
\end{equation}
and both together simplify equations (\ref{s1}) - (\ref{s6}) to
\begin{eqnarray}
&  & \Lambda_0=0                                         \label{f3} \\  
& & 4\Lambda_{5x} - \Lambda_7 =0                        \label{f4} \\  
& & 12\Lambda_{5t} +2\Lambda_5^{\;\;2}+3\Lambda_6 =0    \label{f5} \\  
& & 9p_{xx} - 3p_x\Lambda_4 + p\Lambda_8 =0             \label{f6} \\
& & 9h_{xx} - 3h_x\Lambda_4 + h\Lambda_8=0.              \label{f7} 
\end{eqnarray}
Substitution of $h_t, p_t$ into the integrability condition (\ref{i0}) gives
\[12\Lambda_{4t} - 12\Lambda_{5x} - \Lambda_7=0 \]
which when simplified with equation (\ref{f4}) yields
\begin{equation}
3\Lambda_{4t} - 4\Lambda_{5x}=0. \label{i4}
\end{equation}
The remaining integrability condition between equations (\ref{f1}) and
(\ref{f6}) (and equally (\ref{f2}) and (\ref{f7})) is computed by the 
differentiation of (\ref{f6}) with respect to $t$ and the replacement of 
$p_{txx}, p_{tx}, p_t$ with (\ref{f1}), $p_{xx}$ with (\ref{f6})
and $\Lambda_{4t}$ with (\ref{f4}). The result is
\begin{equation}
6\Lambda_{5xx} - 2\Lambda_{5x}\Lambda_4 + 3\Lambda_{8t}=0. \label{i5}
\end{equation}

\strut\hfill

\noindent
{\bf We summarize:}
The procedure for a given set of expressions $\Lambda_0, \Lambda_1,
\ldots, \Lambda_5$ is as follows.
\begin{enumerate}
\item Compute $\Lambda_6, \Lambda_7, \Lambda_8$ from equations
      (\ref{l1sub}), (\ref{l2sub}), (\ref{l3sub}).
\item The following set of conditions for $\Lambda_i$ ($i=0,\ldots, 5$)
is necessary
      and sufficient for a solution with $F_t=0$, $h_x \neq 0$
      to exist:
\[\Lambda_0=0, \;\;\; 4\Lambda_{5x} - \Lambda_7=0, \;\;\;
  12\Lambda_{5t} +2\Lambda_5^{2}+3\Lambda_6=0, \]
\[3\Lambda_{4t} - 4\Lambda_{5x}=0, \;\;\;
  6\Lambda_{5xx} - 2\Lambda_{5x}\Lambda_4 + 3\Lambda_{8t}=0 .\]
\item A function $u(x,t)$ is to be computed as
      \[u(x,t) = \exp\left[\frac{2}{9}\int \Lambda_5(x,t)\, dt\right].\]
\item The functions $p(t,x)$ and $h(t,x)$ are computed from
      \[p(x,t)=v(x)u(x,t),\;\;\;\; h(x,t)=w(x)u(x,t), \]
      where $v(x)$ and $w(x)$ are to be computed from the two linear
      second order conditions
\[9(vu)_{xx}-3(vu)_x\Lambda_4+(vu)\Lambda_8=0, \;\;\;
 9(wu)_{xx}-3(wu)_x\Lambda_4+(wu)\Lambda_8=0\]
      which when divided by $u$ are purely $x$-dependent
      and therefore are ordinary differential equations for $v(x)$ and $w(x)$ as
      guaranteed by the integrability conditions above.
\item With the last step we satisfied all conditions (\ref{f1}) -
      (\ref{f7}) and computed $p$ and $h$ which give $G$ and $F$
       through equations
      (\ref{gsub}) and (\ref{fsub}):
      \[G(x,t) = h^{-3/2}, \;\;\; F(x) = ph^{-1}.\]
\end{enumerate}

\subsection{The general conditions for $X'''=0$}

In order to invert the system (\ref{3.2}) in general,
that is to solve $F$ and $G$ from (\ref{3.2}), we need to consider
three different cases. The obvious conditions which apply in all cases
are $p\neq 0,\ h\neq 0,\ F_x G\neq 0$. Recall further that $X'''=0$
is transformed into (\ref{3.1}), \viz
\bg
\dddot{x} +\Lambda_5(x,t)\ddot{x} +\Lambda_4(x,t)\dot{ x}\ddot{x}
+\Lambda_3(x,t)\dot{ x}^3+\Lambda_2(x,t)\dot{x}^2
+\Lambda_1(x,t)\dot{x}
+\Lambda_0(x,t) = 0,
\ed
and $\lam_6,\ \lam_7,\ \lam_8$ are defined in
(\ref{l1sub})-(\ref{l3sub})),
\viz
\bg
\Lambda_6(x,t) & = & -6\Lambda_{5t}+6\Lambda_1-2\Lambda_5^{2}\\
\Lambda_7(x,t) & = & \;\;\;6\Lambda_{4t} - 6\Lambda_2 + 2\Lambda_4\Lambda_5\\
\Lambda_8(x,t) & = & -6\Lambda_{4x}+18\Lambda_3-2\Lambda_4^{2}. 
\ed

\strut\hfill

In providing the general conditions in this subsection we
are not
  able to document each individual step like for the previous
  two subcases. Instead we give only the final result, {\it i.e.}\ 
we list the conditions on $h$ and $p$ and the conditions on the
$\Lambda_i$s which must be satisfied for a given third order
ordinary differential equation in order to establish
linearisation to $X'''(T)=0$ by (\ref{1.7}).

Below we use the notation
\bg
\{A(h)=0\}_{h\leftrightarrow p}:= \left\{A(h)=0\
\mbox{and}\ A(h)|_{h\rightarrow
p}=0\right\},
\ed
where $A(h)$ denotes the differential expression in $h$. Thus
$\{A\}_{h\leftrightarrow p}$ represents two
differential expression; one in $h$ and the same differential
expression but with $h$ replaced by $p$.

\strut\hfill

\noindent
{\bf Case I.} $\Lambda_8\neq 0,\ -ph_x+p_xh\neq 0,\
2\Lambda_8\Lambda_4-3\Lambda_{8x}\neq 0$:

\strut\hfill

\noindent
The functions $h$ and $p$ for the transformation
\bg
X(T)=p(x,t)h^{-1}(x,t),\qquad dT=h^{-3/2}(x,t) dt
\ed
are to be solved from the following
set of linear equations:
\bg
& &
\left\{h_{xx}=\frac{1}{3}h_x\lam_4-\frac{1}{9}h\lam_8\right\}_{h\leftrightarrow
p}\\
& & \left\{h_t=\frac{1}{9\lam_8^2}
\left(108h_x\lam_{4t}\lam_8
-108h_x\lam_{5x}\lam_8-9h_x\lam_7\lam_8+2h\lam_8^2\lam_5
-2h\lam_8\lam_{7x}
\right.\right.\\
& &\quad +16h\lam_8\lam_{5x}\lam_4
-4h\lam_8\lam_{8t}
-16h\lam_8\lam_{4t}\lam_4
-2h\lam_7\lam_{8x}
-24 h\lam_{5x}\lam_{8x}\\
& &\quad\left.
\vphantom{\frac{1}{9\lam_8^2}}
\left.\vphantom{frac{1}{3^2}}
+24 h\lam_{4t}\lam_{8x}
+2h\lam_8\lam_7\lam_4\right)
\right\}_{h\leftrightarrow p}.
\ed
The following conditions on $\lam_i$ ($i=1,\ldots, 8$) are to be satisfied:
\bg
& & \lam_{4tt}=\frac{1}{12\lam_8^2}\left(-24\lam_{8x}\lam_{5x}^2
-24\lam_{8x}\lam_{4t}^2
-2\lam_{8x}\lam_7\lam_{5x}
+48\lam_{8x}\lam_{4t}\lam_{5x}\right.\\
& &\quad +2\lam_{8x}\lam_7\lam_{4t}
+16\lam_8\lam_{4t}^2\lam_4
-2\lam_8\lam_{4t}\lam_7\lam_4
+16\lam_8\lam_{5x}^2\lam_4
-2\lam_8\lam_{5x}\lam_{7x}\\
& & \quad
-4\lam_8^2\lam_5\lam_{4t}
-\lam_8^2\lam_{6x}
+4\lam_8^2\lam_5\lam_{5x}
+2\lam_8\lam_{4t}\lam_{7x}
+2\lam_8\lam_{5x}\lam_7\lam_4\\
& & \quad\left.
-16\lam_8\lam_{8t}\lam_{5x}
-32\lam_8\lam_{5x}\lam_{4t}\lam_4
-\lam_8\lam_{8t}\lam_7
+12\lam_8^2\lam_{5xt}
+16\lam_8\lam_{8t}\lam_{4t}\right)\\
& & \\
& & \lam_{5tt}=\frac{1}{72\lam_8^3}\left(
-144\lam_{8x}\lam_{5x}\lam_7\lam_{4t}
+38 \lam_8\lam_7^2\lam_4\lam_{5x}
-96\lam_8\lam_7\lam_4\lam_{5x}^2\right.\\
& & \quad+6\lam_8\lam_{7x}\lam_{4t}\lam_{7}
-96\lam_8\lam_{4t}^2\lam_4\lam_7
-288\lam_8\lam_{7x}\lam_{4t}\lam_{5x}
-6\lam_8\lam_{7x}\lam_{7}\lam_{5x}\\
& & \quad
-84\lam_8^2\lam_5\lam_7\lam_{4t}
-3456\lam_8\lam_{4t}^2\lam_4\lam_{5x}
+60\lam_8\lam_7\lam_{5x}\lam_{8t}
-1440\lam_8\lam_{4t}\lam_{5x}\lam_{8t}\\
& & \quad
-38\lam_8\lam_7^2\lam_4\lam_{4t}
+84\lam_8^2\lam_7\lam_5\lam_{5x}
-60\lam_8\lam_7\lam_{4t}\lam_{8t}
+192\lam_8\lam_7\lam_4\lam_{4t}\lam_{5x}\\
& &\quad
+3456\lam_8\lam_{5x}^2\lam_4\lam_{4t}
-54\lam_{8x}\lam_7^2\lam_{5x}
+5184\lam_{8x}\lam_{5x}\lam_{4t}^2
+72\lam_{8x}\lam_{5x}^2\lam_7\\
& &\quad
+72\lam_{8x}\lam_{4t}^2\lam_7
+144\lam_8\lam_{4t}^2\lam_{7x}
+1152\lam_8\lam_{4t}^3\lam_4
+12\lam_8^2\lam_7^2\lam_5
+144\lam_8\lam_{7x}\lam_{5x}^2\\
& &\quad
-252\lam_8^2\lam_{4t}\lam_{7t}
-216\lam_{4t}\lam_8^2\lam_{6x}
+33\lam_7\lam_8^2\lam_{6x}
-1152\lam_8\lam_{5x}^3\lam_4
-12\lam_8^3\lam_6\lam_5\\
& &\quad
+72\lam_0\lam_8^3\lam_4
-72\lam_8^3\lam_5\lam_{5t}
-5184\lam_{8x}\lam_{4t}\lam_{5x}^2
-5\lam_8\lam_7^2\lam_{8t}
+720\lam_8\lam_{5x}^2\lam_{8t}\\
& &\quad
+36\lam_8^2\lam_{7t}\lam_7
+4\lam_8\lam_7^3\lam_4
+54\lam_{8x}\lam_{4t}\lam_7^2
+216\lam_{5x}\lam_8^2\lam_{6x}
+720\lam_8\lam_{4t}^2\lam_{8t}\\
& &\quad
-4\lam_8\lam_{7x}\lam_7^2
+252\lam_8^2\lam_{7t}\lam_{5x}
-4\lam_{8x}\lam_7^3
-1728\lam_{8x}\lam_{4t}^3
+216\lam_8^3\lam_{0x}\\
& &\quad\left.
-18\lam_8^3\lam_{6t}
+1728\lam_{8x}\lam_{5x}^3
-8\lam_8^3\lam_5^3\,\right)\\
& & \\
& & \lam_{7tt}=\frac{1}{18\lam_8^3}\left(
2\lam_8^2\lam_7\lam_5\lam_{7x}
+252\lam_8^2\lam_{5x}\lam_7\lam_{4t}
+3\lam_7^2\lam_{8t}\lam_8\lam_4
-3\lam_8\lam_{7x}\lam_7\lam_{8t}\right.\\
& & \quad
-3\lam_8^2\lam_{6x}\lam_7\lam_4
-36\lam_{8x}\lam_8\lam_{6x}\lam_{4t}
+48\lam_8^2\lam_{4t}\lam_4\lam_{7t}
+10\lam_8^2\lam_{8t}\lam_5\lam_7\\
& & \quad
-6\lam_8^2\lam_{7t}\lam_4\lam_7
+36\lam_{8x}\lam_8\lam_{6x}\lam_{5x}
+36\lam_{8x}\lam_7\lam_{4t}\lam_{8t}
+6\lam_{8x}\lam_8\lam_{7t}\lam_7\\
& & \quad
+3\lam_{8x}\lam_8\lam_{6x}\lam_7
-24\lam_8^2\lam_{6x}\lam_{5x}\lam_4
+24\lam_8^2\lam_{6x}\lam_{4t}\lam_4
+2\lam_{8x}\lam_8\lam_7^2\lam_5\\
& & \quad
+72\lam_{8x}\lam_8\lam_{7t}\lam_{5x}
-72\lam_{8x}\lam_8\lam_{4t}\lam_{7t}
-36\lam_{8x}\lam_7\lam_{5x}\lam_{8t}
-48\lam_8^2\lam_{7t}\lam_4\lam_{5x}\\
& & \quad
-2\lam_8^2\lam_7^2\lam_4\lam_5
+24\lam_7\lam_{5x}\lam_{8t}\lam_8\lam_4
+16\lam_8^2\lam_5\lam_7\lam_{4t}\lam_4
-24\lam_{8x}\lam_8\lam_5\lam_7\lam_{4t}\\
& & \quad
-24\lam_7\lam_{4t}\lam_{8t}\lam_8\lam_4
+24\lam_{8x}\lam_8\lam_7\lam_5\lam_{5x}
-16\lam_8^2\lam_7\lam_5\lam_{5x}\lam_4
-18\lam_8^3\lam_{6xt}\\
& & \quad
-126\lam_8^2\lam_{5x}^2\lam_7
-6\lam_8^3\lam_7\lam_{5t}
+21\lam_8^2\lam_{4t}\lam_7^2
-126\lam_8^2\lam_{4t}^2\lam_7
-648\lam_8^2\lam_{5x}\lam_{4t}^2\\
& & \quad
+648\lam_8^2\lam_{4t}\lam_{5x}^2
+6\lam_8^3\lam_5^2\lam_{4t}
+6\lam_8^2\lam_{7t}\lam_{7x}
+30\lam_8^2\lam_{7t}\lam_{8t}
-6\lam_8^3\lam_{5x}\lam_5^2\\
& & \quad
-9\lam_8^3\lam_6\lam_{5x}
+9\lam_8^3\lam_6\lam_{4t}
-6\lam_8\lam_7\lam_{8t}^2
-21\lam_8^2\lam_7^2\lam_{5x}
-3\lam_{8x}\lam_7^2\lam_{8t}\\
& & \quad
-36\lam_8^3\lam_{5x}\lam_{5t}
-4\lam_8^3\lam_7\lam_5^2
-12\lam_8^3\lam_{6x}\lam_5
-18\lam_8^3\lam_{7t}\lam_5
+24\lam_8^2\lam_{6x}\lam_{8t}\\
& & \quad\left.
+3\lam_8^2\lam_{6x}\lam_{7x}
+36\lam_8^3\lam_{5t}\lam_{4t}
-18\lam_0\lam_8^4
+216\lam_8^2\lam_{4t}^3
-\lam_8^2\lam_7^3
-216\lam_8^2\lam_{5x}^3\right)\\
& & \\
& & \lam_{8tt}=\frac{1}{36\lam_8^3}\left(
9\lam_8^2\lam_{7xx}\lam_7
-108\lam_8^2\lam_{7xx}\lam_{4t}
+1296\lam_8\lam_{5x}^2\lam_{8xx}
+9\lam_8\lam_7^2\lam_{8xx}\right.\\
& & \quad
-16\lam_8^2\lam_{8t}\lam_7\lam_4
+12\lam_8\lam_{8t}\lam_{8x}\lam_7
+108\lam_8^2\lam_{7xx}\lam_{5x}
+24\lam_8\lam_{4t}\lam_4\lam_7\lam_{8x}\\
& & \quad
-24\lam_8\lam_{5x}\lam_4\lam_7\lam_{8x}
-288\lam_8\lam_{5x}\lam_4\lam_{4t}\lam_{8x}
-192\lam_8^2\lam_{5x}\lam_4^2\lam_{4t}
+8\lam_8^3\lam_5\lam_7\lam_4\\
& & \quad
-144\lam_{5x}\lam_{8x}\lam_8\lam_{7x}
+1728\lam_8^2\lam_{4t}\lam_{4x}\lam_{5x}
+12\lam_8^2\lam_{7x}\lam_{4t}\lam_4
+144\lam_8\lam_{4t}^2\lam_4\lam_{8x}\\
& & \quad
-5\lam_7^2\lam_{8x}\lam_8\lam_4
-12\lam_8^2\lam_{7x}\lam_{5x}\lam_4
-5\lam_8^2\lam_{7x}\lam_4\lam_7
+144\lam_8\lam_{5x}^2\lam_4\lam_{8x}\\
& & \quad
+144\lam_{4t}\lam_{8x}\lam_8\lam_{7x}
+180\lam_8^2\lam_{4t}\lam_{4x}\lam_7
-12\lam_8^2\lam_5\lam_7\lam_{8x}
-6\lam_7\lam_{8x}\lam_8\lam_{7x}\\
& & \quad
+40\lam_8^2\lam_7\lam_4^2\lam_{5x}
-180\lam_8^2\lam_7\lam_{4x}\lam_{5x}
-40\lam_8^2\lam_7\lam_4^2\lam_{4t}
-2592\lam_8\lam_{5x}\lam_{8xx}\lam_{4t}\\
& & \quad
-216\lam_8\lam_7\lam_{8xx}\lam_{4t}
+216\lam_8\lam_7\lam_{8xx}\lam_{5x}
+1296\lam_8\lam_{4t}^2\lam_{8xx}
+18\lam_8^3\lam_{6xx}\\
& & \quad
+24\lam_8^3\lam_{7t}\lam_4
-36\lam_8^2\lam_{7t}\lam_{8x}
-864\lam_8^2\lam_{5x}^2\lam_{4x}
+24\lam_{8}^2\lam_{8t}\lam_{7x}
-216\lam_7\lam_{8x}^2\lam_{5x}\\
& & \quad
+1656\lam_8^3\lam_{5x}\lam_{4t}
+159\lam_8^3\lam_{4t}\lam_7
+96\lam_8^2\lam_{5x}^2\lam_4^2
+4\lam_8^2\lam_7^2\lam_4^2
-144\lam_8^3\lam_{5x}\lam_7\\
& & \quad
+3456\lam_{4t}\lam_{8x}^2\lam_{5x}
+216\lam_7\lam_{8x}^2\lam_{4t}
-12\lam_8^3\lam_{8t}\lam_5
+18\lam_8^3\lam_{6x}\lam_4
-36\lam_8^2\lam_{6x}\lam_{8x}\\
& & \quad
+96\lam_8^2\lam_{4t}^2\lam_4^2
-864\lam_8^2\lam_{4t}^2\lam_{4x}
-9\lam_8^2\lam_7^2\lam_{4x}
+6\lam_8^4\lam_6
-1728\lam_{5x}^2\lam_{8x}^2
-6\lam_7^2\lam_{8x}^2\\
& & \quad
+4\lam_8^4\lam_5^2
+24\lam_8^4\lam_{5t}
-6\lam_8^3\lam_7^2
-792\lam_8^3\lam_{5x}^2
+48\lam_8^2\lam_{8t}^2
\left.
-864\lam_8^3\lam_{4t}^2
-1728\lam_{4t}^2\lam_{8x}^2\right)\\
& & \\
& & \lam_{4xt}=\frac{1}{18\lam_8}\left(
2\lam_8\lam_{7x}
+2\lam_8\lam_{5x}\lam_4
+18\lam_8\lam_{5xx}
+\lam_8\lam_{8t}
-2\lam_{8}\lam_{4t}\lam_4
\right.\\
& & \quad\left.
-\lam_7\lam_{8x}
-12\lam_{5x}\lam_{8x}
+12\lam_{4t}\lam_{8x}\right)\\
& & \\
& & \lam_{7xt}=\frac{1}{36\lam_8^3}\left(
4\lam_8^2\lam_{8t}\lam_7\lam_4
-12\lam_8\lam_{8t}\lam_{8x}\lam_7
-72\lam_8\lam_{4t}\lam_4\lam_7\lam_{8x}
-4\lam_8^3\lam_5\lam_7\lam_4
\right.\\
& &\quad
+72\lam_8\lam_{5x}\lam_4\lam_7\lam_{8x}
+72\lam_{5x}\lam_{8x}\lam_8\lam_{7x}
+48\lam_8^2\lam_{7x}\lam_{4t}\lam_4
+8\lam_7^2\lam_{8x}\lam_8\lam_4\\
& &\quad
-48\lam_8^2\lam_{7x}\lam_{5x}\lam_4
-4\lam_8^2\lam_{7x}\lam_4\lam_7
-72\lam_{4t}\lam_{8x}\lam_8\lam_{7x}
+12\lam_8^2\lam_5\lam_7\lam_{8x}
\\
& &\quad
-16\lam_8^2\lam_7\lam_4^2\lam_{5x}
+16\lam_8^2\lam_7\lam_4^2\lam_{4t}
-36\lam_8^3\lam_{6xx}
-12\lam_8^3\lam_{7t}\lam_4
+36\lam_8^2\lam_{7t}\lam_{8x}\\
& &\quad
+12\lam_8^2\lam_{8t}\lam_{7x}
-72\lam_7\lam_{8x}^2\lam_{5x}
-216\lam_8^3\lam_{5x}\lam_{4t}
-66\lam_8^3\lam_{4t}\lam_{7}
-2\lam_8^2\lam_7^2\lam_4^2\\
& &\quad
+54\lam_8^3\lam_{5x}\lam_7
+72\lam_7\lam_{8x}^2\lam_{4t}
-12\lam_{8}^3\lam_{6x}\lam_4
+36\lam_8^2\lam_{6x}\lam_{8x}
-12\lam_8^3\lam_5\lam_{7x}\\
& &\quad
+6\lam_8^2\lam_{7x}^2
-3\lam_8^4\lam_6
-6\lam_7^2\lam_{8x}^2
-2\lam_{8}^4\lam_5^2
-12\lam_8^4\lam_{5t}
+3\lam_8^3\lam_7^2\\
& &\quad\left.
+72\lam_8^3\lam_{5x}^2
+144\lam_8^3\lam_{4t}^2\right)\\
& & \\
& & \lam_{8xt}=-\frac{1}{18\lam_8^2}\left(
-8\lam_8^2\lam_{4t}\lam_4^2
-12\lam_7\lam_{8x}^2
-144\lam_{5x}\lam_{8x}^2
+144\lam_{4t}\lam_{8x}^2
\right.\\
& &\quad
+9\lam_8^2\lam_{7xx}
-36\lam_8^3\lam_{5x}
-\lam_8^2\lam_{7x}\lam_4
-9\lam_8^2\lam_7\lam_{4x}
+12\lam_8\lam_{8x}\lam_{5x}\lam_4\\
& &\quad
-12\lam_8\lam_{8x}\lam_{4t}\lam_4
+5\lam_8\lam_{8x}\lam_7\lam_4
+8\lam_8^2\lam_{5x}\lam_4^2
+4\lam_8^2\lam_{8t}\lam_4
+9\lam_8\lam_7\lam_{8xx}\\
& &\quad
-24\lam_8\lam_{8t}\lam_{8x}
-108\lam_8\lam_{4t}\lam_{8xx}
+108\lam_8\lam_{5x}\lam_{8xx}
+27\lam_8^3\lam_{4t}\\
& &\quad\left.
-12\lam_{8x}\lam_{8}\lam_{7x}
+72\lam_8^2\lam_{4t}\lam_{4x}
-72\lam_{8}^2\lam_{5x}\lam_{4x}\right).
\ed

\noindent
{\bf Case II.} $\lam_8\neq 0,\ -ph_x+hp_x\neq 0,\
2\lam_8\lam_4-3\lam_{8x}=0$:

\strut\hfill

\noindent
The  functions $h$ and $p$ for the transformation
\bg
X(T)=p(x,t)h^{-1}(x,t),\qquad dT=h^{-3/2}(x,t) dt
\ed
are to be solved from the following
set of linear equations:
\bg
& &\left\{ h_{xx}=\frac{1}{3}h_x\lam_4
-\frac{1}{9}h\lam_8\right\}_{h\leftrightarrow p}\\
& & \left\{h_t=\frac{1}{27\lam_8}
\left(324h_x\lam_{4t}
-324h_x\lam_{5x}-27h_x\lam_7
+6h\lam_8\lam_5\right.\right.\\
& & \quad\left.
\vphantom{\frac{1}{27\lam_8}}
\left.
-6h\lam_{7x}
-12h\lam_{8t}
+2h\lam_7\lam_4\right)\right\}_{h\leftrightarrow p}.
\ed
The following conditions on $\lam_i$ ($i=1,\ldots, 8$) are to be satisfied:
\bg
& & \lam_{6xxx}=
-\frac{13}{27}\lam_7\lam_{8t}
-\frac{4}{27}\lam_{5x}\lam_7\lam_4
-\frac{22}{9}\lam_{7x}\lam_{4t}
+\frac{2}{3}\lam_{5x}^2\lam_4
+\frac{10}{9}\lam_{5x}\lam_{7x}\\
& & \quad
+2\lam_{4t}\lam_{5xx}
-\frac{2}{27}\lam_7\lam_{7x}
-2\lam_{5x}\lam_{5xx}
+\frac{1}{3}\lam_{4x}\lam_{6x}
+\lam_4\lam_{6xx}
-\frac{2}{3}\lam_7\lam_{5xx}\\
& & \quad
-\frac{4}{9}\lam_8\lam_{6x}
-\frac{2}{9}\lam_4^2\lam_{6x}
+\frac{82}{9}\lam_{8t}\lam_{4t}
-\frac{79}{9}\lam_{8t}\lam_{5x}
+\frac{22}{27}\lam_7\lam_4\lam_{4t}\\
& & \quad
+\frac{2}{81}\lam_7^2\lam_4
-\frac{2}{3}\lam_{4t}\lam_{5x}\lam_4\\
& & \\
& & \lam_{4tt}=\frac{1}{36\lam_8}\left(
-48\lam_{8t}\lam_{5x}
+12\lam_8\lam_5\lam_{5x}
-6\lam_{5x}\lam_{7x}
+6\lam_{7x}\lam_{4t}
-3\lam_8\lam_{6x}
\right.\\
& & \quad\left.
+36\lam_8\lam_{5xt}
-2\lam_7\lam_4\lam_{4t}
-3\lam_7\lam_{8t}
-12\lam_8\lam_5\lam_{4t}
+48\lam_{8t}\lam_{4t}
+2\lam_{5x}\lam_7\lam_4\right)\\
& & \\
& &
\lam_{5tt}=\frac{1}{216\lam_8^2}\left(
288\lam_7\lam_4\lam_{4t}\lam_{5x}
+252\lam_8\lam_7\lam_{5x}\lam_5
+99\lam_8\lam_{6x}\lam_7\right.\\
& & \quad
-648\lam_8\lam_{6x}\lam_{4t}
-252\lam_8\lam_5\lam_7\lam_{4t}
+6\lam_7^2\lam_4\lam_{5x}
-144\lam_7\lam_4\lam_{5x}^2
-144\lam_{4t}^2\lam_4\lam_7\\
& & \quad
+648\lam_8\lam_{6x}\lam_{5x}
-36\lam_8^2\lam_6\lam_5
-864\lam_{7x}\lam_{4t}\lam_{5x}
-18\lam_{7x}\lam_7\lam_{5x}
+36\lam_8\lam_7^2\lam_5\\
& & \quad
+216\lam_0\lam_8^2\lam_4
-6\lam_7^2\lam_4\lam_{4t}
-216\lam_8^2\lam_5\lam_{5t}
+180\lam_{8t}\lam_7\lam_{5x}
-180\lam_{8t}\lam_7\lam_{4t}\\
& & \quad
-756\lam_8\lam_{7t}\lam_{4t}
+18\lam_{7x}\lam_{4t}\lam_7
+108\lam_8\lam_7\lam_{7t}
-4320\lam_{8t}\lam_{4t}\lam_{5x}
-54\lam_8^2\lam_{6t}\\
& & \quad
+756\lam_8\lam_{7t}\lam_{5x}
+2160\lam_{8t}\lam_{4t}^2
+648\lam_8^2\lam_{0x}
-12\lam_{7x}\lam_7^2
+4\lam_7^3\lam_4
+432\lam_{4t}^2\lam_{7x}\\
& & \quad\left.
+432\lam_{7x}\lam_{5x}^2
-24\lam_8^2\lam_5^3
+2160\lam_{8t}\lam_{5x}^2
-15\lam_{8t}\lam_7^2\right)\\
& & \\
& &
\lam_{7tt}=-\frac{1}{54\lam_8^2}\left(
6\lam_{8}\lam_{7t}\lam_4\lam_7
+2\lam_8\lam_7^2\lam_4\lam_5
-30\lam_8\lam_5\lam_7\lam_{8t}
+54\lam_8^2\lam_{6xt}\right.\\
& & \quad
-756\lam_8\lam_{4t}\lam_{5x}\lam_7
-6\lam_8\lam_{7x}\lam_5\lam_7
+3\lam_8\lam_{6x}\lam_7\lam_4
+108\lam_8^2\lam_{5x}\lam_{5t}\\
& & \quad
+63\lam_8\lam_{7}^2\lam_{5x}
+18\lam_8^2\lam_7\lam_{5t}
-18\lam_8\lam_{7x}\lam_{7t}
+18\lam_8^2\lam_{5x}\lam_5^2
-63\lam_8\lam_{4t}\lam_7^2\\
& & \quad
+378\lam_8\lam_{4t}^2\lam_7
+1944\lam_8\lam_{5x}\lam_{4t}^2
+9\lam_{7x}\lam_7\lam_{8t}
-3\lam_7^2\lam_{8t}\lam_4
+54\lam_8^2\lam_{7t}\lam_5\\
& & \quad
+378\lam_8\lam_{5x}^2\lam_7
-108\lam_8^2\lam_{5t}\lam_{4t}
-90\lam_8\lam_{7t}\lam_{8t}
+12\lam_8^2\lam_7\lam_5^2
-1944\lam_8\lam_{4t}\lam_{5x}^2\\
& & \quad
-72\lam_8\lam_{6x}\lam_{8t}
+36\lam_8^2\lam_{6x}\lam_5
-9\lam_8\lam_{6x}\lam_{7x}
+27\lam_{5x}\lam_8^2\lam_6
-27\lam_{4t}\lam_8^2\lam_6\\
& & \quad\left.
-18\lam_8^2\lam_5^2\lam_{4t}
+648\lam_8\lam_{5x}^3
+54\lam_0\lam_8^3
+3\lam_8\lam_7^3
-648\lam_8\lam_{4t}^3
+18\lam_7\lam_{8t}^2\right)\\
& & \\
& &
\lam_{8tt}=\frac{1}{18\lam_8}\left(
24\lam_{8t}^2
-6\lam_8\lam_{8t}\lam_5
+60\lam_8\lam_{4t}\lam_7
-54\lam_8\lam_7\lam_{5x}
+378\lam_8\lam_{4t}\lam_{5x}
\right.\\
& & \quad
-4\lam_{8t}\lam_7\lam_4
-3\lam_8\lam_7^2
+12\lam_8^2\lam_{5t}
+2\lam_8^2\lam_5^2
+12\lam_{8t}\lam_{7x}
+3\lam_8^2\lam_6
-198\lam_8\lam_{4t}^2\\
& & \quad\left.
-180\lam_8\lam_{5x}^2
+9\lam_8\lam_{6xx}
-3\lam_4\lam_8\lam_{6x}\right)\\
& & \\
& & \lam_{4xt}=\frac{1}{9}\lam_{7x}
-\frac{1}{3}\lam_{5x}\lam_4
+\lam_{5xx}
+\frac{1}{18}\lam_{8t}
+\frac{1}{3}\lam_{4t}\lam_4
-\frac{1}{27}\lam_7\lam_4\\
& & \\
& & \lam_{7xt}=-\frac{1}{108\lam_8}\left(
12\lam_{7x}\lam_4\lam_7
+36\lam_8\lam_5\lam_{7x}
-162\lam_8\lam_7\lam_{5x}
+648\lam_8\lam_{4t}\lam_{5x}\right.\\
& &\quad
-12\lam_8\lam_5\lam_7\lam_4
-2\lam_7^2\lam_4^2
+9\lam_8^2\lam_6
+6\lam_8^2\lam_5^2
-18\lam_{7x}^2
+36\lam_8^2\lam_{5t}
-9\lam_8\lam_7^2\\
& &\quad
+12\lam_{8t}\lam_7\lam_4
-36\lam_4\lam_8\lam_{6x}
+108\lam_8\lam_{6xx}
-216\lam_{8}\lam_{5x}^2
-36\lam_{8t}\lam_{7x}\\
& &\quad\left.
-432\lam_8\lam_{4t}^2
-36\lam_8\lam_{7t}\lam_4
+198\lam_8\lam_{4t}\lam_7\right)\\
& & \\
& & \lam_{7xx}=-\frac{2}{9}\lam_7\lam_4^2
-\frac{13}{3}\lam_8\lam_{4t}
+4\lam_8\lam_{5x}
+\lam_{7x}\lam_4
+\frac{1}{3}\lam_7\lam_{4x}.
\ed

\noindent
{\bf Case III.} $\lam_8= 0,\ \lam_7\neq 0,\
-ph_x+hp_x\neq 0,\quad p_x\neq 0$:

\strut\hfill

\noindent
The  functions $h$ and $p$ in the transformation
\bg
X(T)=p(x,t)h^{-1}(x,t),\qquad dT=h^{-3/2}(x,t) dt
\ed
are to be solved from the following
set of linear equations:
\bg
& &\left\{ h_{xx}=\frac{1}{3}h_x\lam_4\right\}_{h\leftrightarrow p}\\
& &\\
& & \left\{ h_t=\frac{1}{9\lam_7^3}\left(
-12h\lam_{6x}\lam_7^2
-12h\lam_{7t}\lam_7^2
-2h\lam_5\lam_7^3
-1296h_x\lam_{7t}\lam_{6x}
-432h_x\lam_{6x}^2\right.\right.\\
& & \quad
+72h_x\lam_7^2\lam_{5t}
-864h_x\lam_{7t}^2
-144h_x\lam_7\lam_{6x}\lam_5
-144h_x\lam_7\lam_5\lam_{7t}
+432h_x\lam_7\lam_{6xt}\\
& & \quad\left.
\vphantom{\frac{1}{9\lam_7^3}}\left.
+432h_x\lam_7\lam_{7tt}
-12h_x\lam_7^2\lam_5^2
-18h_x\lam_7^2\lam_6\right)\right\}_{h\leftrightarrow p}.
\ed
The following conditions on $\lam_i$ ($i=1,\ldots, 8$) are to be satisfied:
\bg
& & \lam_{5ttt}=\frac{1}{\lam_7^4}\left(
-\frac{13}{9}\lam_7^3\lam_{5t}\lam_{7tt}
+\frac{83}{270}\lam_7^3\lam_5^2\lam_{7tt}
-\frac{10}{9}\lam_7^3\lam_{6xt}\lam_{5t}
-\frac{38}{3}\lam_7^2\lam_{7tt}\lam_{6xt}\right.\\
& & \quad
+\frac{136}{15}\lam_7\lam_{6xt}\lam_{6x}^2
+\frac{404}{15}\lam_7\lam_{7t}^2\lam_{7tt}
+\frac{31}{36}\lam_{7}^3\lam_6\lam_{7tt}
+\frac{34}{135}\lam_7^3\lam_{6xt}\lam_5^2\\
& & \quad
+\frac{7}{9}\lam_7^3\lam_{6xt}\lam_6
+\frac{166}{15}\lam_7\lam_{7tt}\lam_{6x}^2
+\frac{344}{15}\lam_7\lam_{7t}^2\lam_{6xt}
-\frac{6}{5}\lam_7^4\lam_{5tt}\lam_5\\
& & \quad
+\frac{12}{5}\lam_7^3\lam_{5tt}\lam_{7t}
+\frac{6}{5}\lam_7^3\lam_{6x}\lam_{5tt}
-\frac{22}{3}\lam_7^2\lam_{7tt}^2
-\frac{1}{4}\lam_7^4\lam_{6tt}
-\frac{16}{3}\lam_7^2\lam_{6xt}^2\\
& & \quad
+3\lam_7^4\lam_{0xt}
-64\lam_{7t}^3\lam_{6x}
-\frac{1}{8}\lam_7^5\lam_0
-\frac{5}{216}\lam_7^4\lam_6^2
-\frac{908}{15}\lam_{7t}^2\lam_{6x}^2
-\frac{61}{2430}\lam_7^4\lam_5^4\\
& & \quad
-\frac{28}{27}\lam_7^4\lam_{5t}^2
-\frac{124}{5}\lam_{6x}^3\lam_{7t}
+\frac{38}{45}\lam_{7}^2\lam_{5t}\lam_{6x}^2
-\frac{43}{810}\lam_7^4\lam_5^2\lam_6
+\frac{112}{45}\lam_7^2\lam_{7t}^2\lam_{5t}\\
& & \quad
+\frac{44}{15}\lam_7^2\lam_{7t}\lam_{6x}\lam_{5t}
-\frac{6}{5}\lam_7^3\lam_0\lam_4\lam_{6x}
-\frac{12}{5}\lam_7^3\lam_0\lam_4\lam_{7t}
+\frac{194}{135}\lam_7^3\lam_{5t}\lam_{6x}\lam_5\\
& & \quad
-\frac{86}{9}\lam_7\lam_{6x}^2\lam_{7t}\lam_5
-\frac{67}{30}\lam_7^2\lam_{6x}\lam_{7t}\lam_6
-\frac{368}{15}\lam_{7t}^4
-\frac{56}{15}\lam_{6x}^4
+\lam_7^4\lam_{0t}\lam_4\\
& & \quad
-\frac{36}{5}\lam_7^3\lam_{0x}\lam_{7t}
+\frac{3}{5}\lam_7^4\lam_{0x}\lam_5
-\frac{18}{5}\lam_7^3\lam_{0x}\lam_{6x}
-\frac{8}{15}\lam_7^2\lam_5^2\lam_{6x}^2
+\lam_7^4\lam_0\lam_{5x}\\
& & \quad
+\frac{1}{5}\lam_7^4\lam_0\lam_4\lam_5
+\frac{97}{540}\lam_7^3\lam_6\lam_{7t}\lam_5
-\frac{7}{270}\lam_7^3\lam_5\lam_{6x}\lam_6
+\frac{353}{135}\lam_7^3\lam_5\lam_{5t}\lam_{7t}\\
& & \quad
-\frac{118}{9}\lam_7\lam_{7t}^2\lam_{6x}\lam_5
-\frac{35}{27}\lam_7^2\lam_{6x}\lam_{7t}\lam_{5}^2
-\frac{52}{9}\lam_7\lam_{7t}^3\lam_5
-\frac{73}{45}\lam_7^2\lam_{7t}^2\lam_6\\
& & \quad
+\frac{3}{5}\lam_7^3\lam_{6t}\lam_{7t}
-\frac{413}{810}\lam_7^4\lam_{5t}\lam_{5}^2
-\frac{7}{108}\lam_7^4\lam_{5t}\lam_6
-\frac{13}{60}\lam_7^4\lam_{6t}\lam_5
-\frac{20}9\lam_7\lam_{6x}^3\lam_5\\
& & \quad
-\frac{32}{45}\lam_7^2\lam_{6x}^2\lam_6
+\frac{3}{10}\lam_{7}^3\lam_{6t}\lam_{6x}
+\frac{169}{810}\lam_7^3\lam_5^3\lam_{7t}
-\frac{112}{135}\lam_7^2\lam_{7t}^2\lam_5^2
+\frac{29}{405}\lam_7^3\lam_{5}^3\lam_{6x}\\
& & \quad
+\frac{118}{45}\lam_7^2\lam_{7t}\lam_{6xt}\lam_5
+\frac{148}{45}\lam_7^2\lam_{7t}\lam_5\lam_{7tt}
+\frac{148}{5}\lam_7\lam_{7t}\lam_{6x}\lam_{6xt}\\
& & \quad\left.
+\frac{178}{5}\lam_7\lam_{7t}\lam_{7tt}\lam_{6x}
+\frac{124}{45}\lam_7^2\lam_{6xt}\lam_{6x}\lam_5
+\frac{154}{45}\lam_7^2\lam_{7tt}\lam_{6x}\lam_5\right)\\
& & \\
& & \lam_{7ttt}=-\frac{1}{\lam_7^2}\left(
-\frac{14}{5}\lam_7\lam_{6x}\lam_{6xt}
-\frac{1}{5}\lam_7^2\lam_{6xt}\lam_5
-\frac{33}{5}\lam_{7x}\lam_{7t}\lam_{7tt}
-\frac{19}{5}\lam_7\lam_{7tt}\lam_{6x}
\right.\\
& & \quad
-\frac{28}{5}\lam_7\lam_{7t}\lam_{6xt}
-\frac{1}{5}\lam_7^2\lam_5\lam_{7tt}
+\frac{4}{15}\lam_7^3\lam_{5tt}
+\lam_7^2\lam_{6xtt}
+6\lam_{7t}\lam_{6x}^2
-\frac{3}{10}\lam_7^3\lam_{0x}\\
& & \quad
-\frac{1}{60}\lam_7^3\lam_{6t}
+\frac{62}{5}\lam_{7t}^2\lam_{6x}
+\frac{4}{5}\lam_{6x}^3
+\frac{36}{5}\lam_{7t}^3
+\frac{1}{90}\lam_7^3\lam_6\lam_5
+\frac{1}{15}\lam_7^3\lam_{5t}\lam_5\\
& & \quad
-\frac{1}{10}\lam_7^3\lam_0\lam_4
-\frac{1}{45}\lam_7^2\lam_{5}^2\lam_{6x}
+\frac{3}{5}\lam_7\lam_{7t}^2\lam_5
+\frac{2}{15}\lam_7\lam_5\lam_{6x}^2
+\frac{11}{15}\lam_7\lam_{7t}\lam_{6x}\lam_5\\
& & \quad\left.
+\frac{1}{135}\lam_7^3\lam_5^3
+\frac{1}{30}\lam_7^2\lam_{6x}\lam_6
-\frac{7}{15}\lam_7^2\lam_{6x}\lam_{5t}
+\frac{1}{15}\lam_{7}^2\lam_6\lam_{7t}
-\frac{3}{5}\lam_7^2\lam_{7t}\lam_{5t}\right)\\
& & \\
& & \lam_{5xtt}=\frac{1}{\lam_7}\left(
\frac{4}{3}\lam_{7t}\lam_{6x}
+\frac{1}{216}\lam_7^2\lam_6
-\frac{5}{324}\lam_7^2\lam_5^2
-\frac{17}{108}\lam_{7}^2\lam_{5t}
-\frac{7}{9}\lam_7\lam_{6xt}\right.\\
& & \quad
-\frac{19}{36}\lam_7\lam_{7tt}
+\frac{8}{9}\lam_{7t}^2
+\frac{4}{9}\lam_{6x}^2
-\frac{2}{27}\lam_7\lam_{6x}\lam_5
+\frac{7}{108}\lam_7\lam_5\lam_{7t}
+\lam_7\lam_0\lam_{4x}\\
& & \quad\left.
-\lam_7\lam_{5t}\lam_{5x}
+3\lam_7\lam_{0xx}
-\frac{1}{6}\lam_7\lam_6\lam_{5x}
-\lam_7\lam_{5}\lam_{5xt}
-\frac{1}{3}\lam_7\lam_{5}^2\lam_{5x}
+\lam_7\lam_4\lam_{0x}\right)\\
& & \\
& & \lam_{6xx}=-\frac{1}{2}\lam_7\lam_{5x}
-\frac{5}{72}\lam_7^2
+\frac{1}{3}\lam_4\lam_{6x}\\
& & \\
& & \lam_{4t}=\lam_{5x}
+\frac{1}{12}\lam_7\\
& & \\
& &\lam_{7x}=\frac{1}{3}\lam_7\lam_4.
\ed

\noindent
{\it Remark:} {\it From our general calculations two more
cases were derived which are not listed in this subsection,
namely the case
$\Lambda_7=0$, $\Lambda_8=0$, $\Lambda_{6x}=0$,
$-ph_x+hp_x\neq 0$, $p_x\neq 0$ and the case
$\Lambda_7=0$, $\Lambda_8=0$, $\Lambda_{6x}=0$,
$-ph_x+hp_x\neq 0$, $p_x= 0$. Both these cases
provide identical conditions on the $\Lambda_i$s
already given in subsection 2.1., {\it i.e.}\
for $G_x=0$. This means that for 
  these ordinary differential equations (specified by these
$\Lambda_i$s)
we already have a simpler
linearisation procedure with $p_x=0$. Recall that our aim is not
to provide all linearisation procedures for a given ordinary
differential equation but rather to find all linearisable
equations allowed by (\ref{1.7}). }


\strut\hfill

We end this subsection with
an example of a third order ordinary differential equation
which we construct
to be linearisable by the generalised Sundman transformation.
We do not claim any physical relevance for this equation.

\strut\hfill

\noindent
{\bf Example 2.1:} The equation
\beg
\label{Ex2.1}
& & \dddot x + \frac{3xe^x-e^x}{x(1+te^x)}\ddot x
+\frac{3xte^x-4te^x-4}{x(1+te^x)}\dot x\ddot x
+\frac{x^2te^x-3xte^x+3te^x+3}{x^2(1+te^x)}\dot x^3\nonumber\\
& & \nonumber\\
& & \quad +\frac{3x^2e^x-6xe^x+3e^x}{x^2(1+te^x)}\dot x^2=0
\eeq
may be reduced to $X'''=0$ by the transformation
\bg
X(T)=te^x+x,\qquad dT=xdt,
\ed
that is
\bg
h(x,t)=x^{-2/3},\qquad p(x,t)=tx^{-2/3}e^x+x^{1/3}.
\ed
By (\ref{2_first_int}) and the transformation given here
two time dependent first integrals follow
for (\ref{Ex2.1}).
This example corresponds to Case I above.



\subsection{On the computations}

Some remarks about the computations with computer algebra are in order.
The computational part of this section consists in determining all
solutions $F(x,t)$ and $G(x,t)$ of the system (\ref{3.2}), {\it i.e.}\
for each class
of solutions a set of conditions for $\Lambda_0, \ldots, \Lambda_5$
and a list of instructions of how to compute $F$ and $G$. Both are
obtained by investigating integrability conditions, like
$\partial_xF_{txx}=\partial_tF_{xxx}$ 
with $F_{txx}$ and $F_{xxx}$ being replaced by corresponding
expressions from two equations. An algorithm for completing this task 
in a finite and systematic way is the well-known (Pseudo-)Differential
Gr\"{o}bner Basis algorithm. To apply it one has to specify a total
ordering of all partial derivatives of all functions by firstly
specifying a lexicographical ordering between all functions and also
between all variables. In addition one can require that the total
derivative of a function should have a higher priority than
lexicographical ordering which was the case in these
computations.

One of the programs that has this algorithm
implemented is the package {\sc Rif} by Allan Wittkopf and Greg Reid
which was used for the general cases I - III. Its strength is that it
is very efficient, especially for larger nonlinear problems.
It can be downloaded from the World Wide Web site with URL\newline
\verb+http://www.cecm.sfu.ca/~wittkopf/rif.html+.

For the two special cases $G_x=0$ and $F_t=0$ the package {\sc Crack}
of TW was used. With its possibility to record the `history' of any
derived equation, {\it i.e.}\ how it results from other equations,
and with its optional fully interactive mode of operation it provided
a convenient way for simplifying system (\ref{3.2}) to
system (\ref{s1}) - (\ref{s6}).
This made it easily possible to prove all conclusions for these two cases
as one could explicitly describe how each condition is generated.
For the harder more general cases computed by {\sc Rif} tracing the
history of equations would be too costly, so one has to trust 
the computation, although the remaining equations for $F$ and
$G$ and the conditions 
for the $\Lambda_i$ have been tested with several third order linearisable 
ordinary differential equations available to the authors. 
The program {\sc Crack} can be downloaded from
\verb+ http://lie.math.brocku.ca/twolf/home/crack.html+.

\section{A special Sundman transformation}
\setcounter{equation}{0}

\noindent
In this section we consider a Sundman transformation of a special form
and construct several examples of linearisable ordinary differential
equations of second, third and fourth order. Some well-known equations
are shown to be included in this construction. Note that we allow
here linearisation to a more general linear form than $X'''(T)=0$,
whereby
the transformation functions $F$ and $G$ in the generalised Sundman
transformation (\ref{1.7}) are special.

We consider the following special form of the Sundman transformation:
\beg
\label{sundman}
& & X(T(t))=x(t)^p\nonumber\\
& & \\
& & dT(t)=x(t)^ndt,\nonumber
\eeq
where $p,n\in{\cal Q}\backslash\{0\}$.
The first three prolongations are
\bg
& & X'=px^{p-n-1}\dot x\\
& & X''=p(p-n-1)x^{p-2n-2}\dot x^2
+px^{p-2n-1}\ddot x\\
& & X'''=p(p-n-1)(p-2n-2)x^{p-3n-3}\dot x^3
+p(3p-4n-3)x^{p-3n-2}\dot x\ddot x\\
& & \qquad +px^{p-3n-1}\dddot x.
\ed

We consider several examples.

\strut\hfill

\noindent
{\bf Example 3.1}: Let
\beg
\label{2nd-I}
X''=I_0,
\eeq
where $I_0$ is an arbitrary constant. By the Sundman
transformation (\ref{sundman}) we obtain in terms of the
coordinates $(x,t)$ the
following equation:
\beg
\label{2nd-pn}
p(p-n-1)x^{p-2n-2}\dot x^2+px^{p-2n-1}\ddot x=I_0.
\eeq
Equation (\ref{2nd-pn}) can thus be linearised by (\ref{sundman}).
The first integral for (\ref{2nd-I}) is
\bg
I_1(X,X')=X-\frac{1}{2I_0}(X')^2
\ed
so that a first integral for (\ref{2nd-pn}) takes the form
\bg
I_1(x,\dot x)=x^p-\frac{p^2}{2I_0}x^{2p-2n-2}\dot x^2.
\ed
As a {special case} we consider $p=n+1$. Then (\ref{2nd-pn})
simplifies to
\beg
\label{2nd-n+1}
\ddot x=\frac{I_0}{n+1}x^n.
\eeq
Solving (\ref{2nd-n+1}) for $I_0$ and considering this $I_0$ this as a first
integral of the third order equation which results upon differentiation
$I_0$ with respect to $t$, we obtain
\beg
\label{3rd-n}
\dddot x-nx^{-1}\dot x\ddot x=0.
\eeq
Thus (\ref{3rd-n}) can be linearised by
\beg
\label{sund-3rd}
X=x^{n+1},\qquad
dT=x^ndt
\eeq
to the equation
\bg
X'''(T)=0.
\ed
Note that (\ref{3rd-n}) is a special case of the generalised Sundman
transformation (\ref{1.7}) with $F_t=0$, derived in general in
subsection 2.2. Equation (\ref{3rd-n}) admits two first integrals, namely
$I_0$ (solved from (\ref{2nd-n+1})) as well as
\bg
I_1(X,X',X'')=XX''-\frac{1}{2}(X')^2
\ed
expressed in the coordinates $(x,t)$ by (\ref{sund-3rd}).

\strut\hfill

\noindent
{\bf Example 3.2}: Let
\beg
\label{3rd-I}
X'''(T)=I_0,
\eeq
where $I_0$ is an arbitrary constant. By the Sundman
transformation (\ref{sundman}) we obtain in terms of the
coordinates $(x,t)$ the
following equation:
\beg
\label{3rd-pn}
& & p(p-n-1)(p-2n-2)x^{p-3n-3}\dot x^3
+p(3p-4n-3)x^{p-3n-2}\dot x\ddot x\nonumber \\
& & \qquad +px^{p-3n-1}\dddot x=I_0
\eeq
Equation (\ref{3rd-pn}) can thus be linearised by (\ref{sundman}).
Two first integrals for (\ref{3rd-I}) are
\bg
& & I_1(X',X'')=X'-\frac{1}{2I_0}(X'')^2\\
& & I_2(X,X',X'')=X +\frac{1}{3}I_0^{-2}(X'')^3-I_0^{-1}X'X''
\ed
so that two first integrals for (\ref{3rd-pn})
are obtained by expressing the above first integrals in
$(x,t)$ variables
using the corresponding Sundman transformation (\ref{sundman}).

As a {special case} we let $p=-1$ and $n=-3/2$. Equation (\ref{3rd-pn})
then simplifies to
\beg
\label{3rd-5/2}
\dddot x=-I_0x^{-5/2}
\eeq
and it follows that (\ref{3rd-5/2}) can be linearised by
\beg
\label{sundT-5/2}
X=x^{-1},\qquad
dT=x^{-3/2}dt
\eeq
to the equation
\bg
X'''(T)=I_0.
\ed
Solving (\ref{3rd-5/2}) for $I_0$ and
considering $I_0$ as a first
integral of the fourth order equation which results upon differentiation
$I_0$ to $t$, we obtain
\beg
\label{4th-5/2}
2x x^{(4)}+5\dot x \dddot x=0.
\eeq
Equation (\ref{4th-5/2}) plays an important role in the symmetry
classification of the anharmonic oscillator \cite{Leach}.
By the above construction (\ref{4th-5/2}) can be linearised
by (\ref{sundT-5/2})
to the equation
\bg
X^{(4)}=0.
\ed
Equation (\ref{4th-5/2}) admits three first integrals, namely
$I_0$ (solved from (\ref{3rd-5/2})), and
\bg
& & I_1(X',X'',X''')=X'X'''-\frac{1}{2}(X'')^2\\
& & I_2(X,X',X'',X''')=
X\left(X'''\right)^2+\frac{1}{3}(X'')^3-X'X''X'''
\ed
which are to be expressed in the coordinates $(x,t)$ by (\ref{sundT-5/2}).

\strut\hfill

\noindent
{\bf Example 3.3}: We consider
\beg
\label{norikati}
\dddot x-x^{-1}\dot x\ddot x-4ax^2\dot x=0,
\eeq
where $a$ is an arbitrary constant. Equation (\ref{norikati})
is of Rikitake type \cite{Leach 2000}.
By the Sundman transformation
(\ref{sundman}) with $p=2$ and $n=1$, {\it i.e.}
\beg
X=x^2,\qquad dT=xdt,
\eeq
it follows that (\ref{norikati})
linearises to
\beg
X'''-4aX'=0.
\eeq

More examples of linearisable equations using the special Sundman
transformation (\ref{sundman}) are given in \cite{berkovich}.

\section{An extended Sundman transformation}
\setcounter{equation}{0}

\noindent
In this final section we propose a further extension of the
generalised Sundman transformation (\ref{1.7}).

Consider the following exact one-form:
\bg
dT(t,x(t))=G_1(t,x)dt+G_2(t,x,\dot x,\ddot x,\ldots )dx,
\ed
where $G_1$ is a smooth function of $x$ and $t$, whereas
$G_2$ is a smooth function of $x,\ t$ and a finite number of
derivatives
of $x$ with respect to $t$.
Here $d$ is the exterior derivative
and by the Lemma of Poincar\'e it is known that
\bg
d(dT)\equiv 0.
\ed
To find the relations between $G_1$ and $G_2$ we calculate
\beg
& & d(dT(x,t))=dG_1\wedge dt+dG_2\wedge dx\nonumber\\
& & \qquad
=\pde{G_1}{x}dx\wedge dt+
\pde{G_2}{t}dt\wedge dx+\pde{G_2}{\dot x}d\dot x\wedge
dx+\pde{G_2}{\ddot x}d\ddot x\wedge dx+\cdots \nonumber\\
& & \qquad
=\pde{G_1}{x}dx\wedge dt +\pde{G_2}{t}dt\wedge dx
+\pde{G_2}{\dot x}\ddot xdt\wedge dx+\pde{G_2}{\ddot x}\dddot x dt\wedge
dx +\cdots\nonumber \\ 
& & \qquad\equiv 0.
\eeq
\noindent
It follows that
\bg
\pde{G_1}{x}=\pde{G_2}{t}+\ddot x\pde{G_2}{\dot x}+\dddot
x\pde{G_2}{\ddot x}+\cdots.
\ed
Also
\beg
& & \pde{T}{t}=G_1(t,x)\\
& & \pde{T}{x}=G_2(t,x,\dot x,\ddot x,\ldots ).
\eeq

We now propose the transformation
\beg
\label{extend_ST}
& & X(T(t,x))=F(x,t)\nonumber\\
& & dT(t,x)=G_1(t,x)dt+G_2(t,x,\dot x,\ddot x,\ldots )dx,
\eeq
with
\beg
\label{mix}
\pde{G_1}{x}=\pde{G_2}{t}+\ddot x\pde{G_2}{\dot x}+\dddot
x\pde{G_2}{\ddot x}+\cdots
\eeq
\noindent
and name this the 
{\bf extended Sundman transformation}.
Note that with $G_2=0$, (\ref{extend_ST}) takes the form of the generalised
Sundman transformation (\ref{1.7}).

Next we derive the first two prolongations of (\ref{extend_ST}):
\beg
\label{der_1}
& & X'(G_1+\dot x G_2)=F_t+\dot xF_x\\ 
& & X''(G_1+\dot x G_2)^2\nonumber\\
\label{der_3}
& & 
\quad +X'\left[G_{1t}+\dot x G_{1x}+ \dot x\left(G_{2t}+\dot
xG_{2x}+\ddot xG_{2\dot x}+\dddot xG_{2\ddot x}+\cdots\right)
+\ddot x G_2\right]\nonumber\\
& & \quad =F_{tt}+2\dot xF_{xt}+\dot x^2F_{xx}+\ddot xF_{x}.
\eeq
Substituting $G_{2t}$ from (\ref{mix}) in (\ref{der_3}) we obtain
\beg
\label{der_2}
& & X''(G_1+\dot x G_2)^2
+X'\left[G_{1t}+2\dot x G_{1x}+\dot x^2
G_{2x}+\ddot x G_2\right]\nonumber\\
& & \qquad =F_{tt}+2\dot xF_{xt}+\dot x^2F_{xx}+\ddot xF_{x}.
\eeq

\strut\hfill

\noindent
We give two examples.

\strut\hfill

\noindent
{\bf Example 5.1}: Consider the first order ordinary differential equation
\bg
X'=X
\ed
and let
\bg
F(t)=t,\qquad G_1(x,t)=xt
\ed
so that $x(t)=t^{-2}$. By (\ref{der_1}) we obtain
\beg
\label{1st_g_2}
G_2(t,x,\dot x)=\left(\frac{\dot F}{F}-G_1\right)\frac{1}{\dot x}
=\left(\frac{1}{t}-xt\right)\frac{1}{\dot x}.
\eeq
Insert $G_2$ given by (\ref{1st_g_2}) and $G_1=xt$
in (\ref{mix}) to obtain the second order ordinary differential equation
\bg
\left(-\frac{1}{t^2}-x\right)\frac{1}{\dot x}-\frac{\ddot x}{\dot x^2}
\left(\frac{1}{t}-xt\right)=t
\ed
which may be written in the form
\bg
\ddot x\left(\frac{1}{t}-xt\right)+\dot x^2t+\dot
x\left(\frac{1}{t^2}+x\right)=0.
\ed
Obviously $x=t^{-2}$ satisfies this second order ordinary differential
equation.
The transformation is
\bg
x=\left(X'\right)^{-1}e^{-T},\qquad t=e^T.
\ed

\strut\hfill

\noindent
{\bf Example 5.2:} Consider the second order ordinary differential equation
\bg
X''(T)=0
\ed
and let
\bg
F(t)=t,\qquad G_1(x,t)=xt
\ed
so that $x(t)=t^{-1}$. By (\ref{der_2}) we obtain
\beg
\label{ex_2}
x+2\dot x t+\ddot x G_2+\ddot x^2 G_{2x}=0,
\eeq
where $G_2=G_2(t,x,\dot x,\ddot x)$.
We now solve $G_2$ from (\ref{ex_2}) by integration. For example,
a solution is
\beg
\label{g_2}
G_2(t,x,\dot x,\ddot x)=-\frac{x}{\ddot x}
+\left(\frac{\dot x}{\ddot x}\right)^2-2\frac{\dot x t}{\ddot x}.
\eeq
Insertion of $G_2$, given by (\ref{g_2}), and $G_1=xt$ into (\ref{mix})
gives the third order ordinary differential equation
\bg
x\ddot x\dddot x+2t\dot x\ddot x\dddot x-2\dot x^2\dddot x-3t\ddot
x^3=0
\ed
for which $x=t^{-1}$ is a solution. The transformation is
\bg
x=\left(X'\right)^{-1}T^{-1},\qquad t=T.
\ed

\strut\hfill

We note that the linearisations which follow from the
extended Sundman transformation (\ref{extend_ST})
given in the examples above
are `downwards', {\it i.e.}\ to a lower order ordinary
differential equation.

\section*{Acknowledgements}

PGLL and TW thank the Department of Mathematics, Lulea University of
Technology, for its kind hospitality while this work was initiated
and PGLL acknowledges the continued support of the National Research
Foundation of South Africa and the University of Natal.
TW thanks Mark Hickman for comments regarding the program {\sc Rif}.
ME acknowledges financial support from the Knut and Alice Wallenberg
Foundation under grant Dnr. KAW 2000.0048.

\end{document}